\documentclass{ws-procs9x6}            
\begin{document}
\title{Orbital stability of the restricted three body problem in General Relativity}

\author{M. Abishev$^1$, H. Quevedo$^2$, S. Toktarbay$^{1*}$ and B. Zhami$^1$}

\address{Physical-Technical Faculty, Al-Farabi Kazakh National University,\\
Al Farabi av. 71, 050040 Almaty, Kazakhstan\\
 Instituto de Ciencias Nucleares, 
Universidad Nacional Aut\'onoma de M\'exico, \\
 AP 70543, M\'exico, DF 04510, Mexico\\ 
$^*$E-mail: saken.yan@yandex.com}

\begin{abstract}
We consider the problem of orbital stability of the motion of a test particle in the restricted three-body problem, by using the orbital moment and its time derivative. We show that it is possible to get some insight into the stability properties of the motion of test particles, without knowing the exact solutions of the motion equations. 
\end{abstract}


\bodymatter

\ 

The relativistic equation of translational motion of a test body in the field of two bodies in general relativity (GR) mechanics, corresponding to the circular restricted three-body problem was investigated in Ref.\citenum{MAthree}. The evolution equations of motion can be studied by using the asymptotic methods of adiabatic theory, according to which the vector elements $\vec{M}$ (orbital moment) and $\vec{A}$ (Laplace vector) provides information about the stability properties of a test body orbit\cite{book1, book2}. 

If we assume that the relativistic corrections to the motion of a test particle satisfy following limits  
\begin{eqnarray} \label{rel}
U_{1} <<c^{2},\ \ U_{2} <<U_{1},\ \ U_{1} /U_{2} \approx v^{2} /c^{2},
\end{eqnarray}
where $U_{1}$ and $U_{2} $ are the potentials of the central and second body, respectively.

The Hamilton function of the system can be written as \cite{MAthree}
\begin{equation}
H=H_{0}+H_{rel},
\end{equation} \label{H}   
where $H_{0}$ and $H_{rel}$ are the classical and relativistic Hamilton functions. We can find the evolution equations of motion of the test particle (third body), which describes the average change of its orbital momentum 
\begin{eqnarray} \label{formula M}
\vec{M}=\left[\vec{r}_{3} ,\vec{p}_{3} \right]=\vec{M}_{0}+\vec{M}_{rel},
\end{eqnarray}
and its time derivative
\begin{eqnarray} \label{formula dotM}
\dot{\vec{M}}=\left[\dot{\vec{r}}_{3} ,\vec{p}_{3} \right]+\left[\vec{r}_{3} ,\dot{\vec{p}}_{3} \right]=\dot{\vec{M}}_{0}+\dot{\vec{M}}_{rel},
\end{eqnarray}
where
\begin{eqnarray} \label{formula dotr3 dotp3}
\dot{\vec{r}}_{3} =\frac{\partial H}{\partial \vec{p}_{3} }, \ \
\dot{\vec{p}}_{3} =-\frac{\partial H}{\partial \vec{r}_{3} }.
\end{eqnarray}
In order to obtain the evolution equations of motion one needs to integrate the Eq. \eqref{formula dotM} for a repetition period of the system configurations T (synodic period of the test body):
\begin{equation} \label{intdotM}
\overline{\dot{\vec{M}}}=\frac{1}{T} \int _{0}^{T}\dot{\vec{M}}dt.
\end{equation}

It is clear that if the test particle is moving on a circular orbit, then the integral  (\ref{intdotM}) is equal to zero \cite{MAthree}.
If the particle is moving in nearly circular orbit with small values of the eccentricity, the equations of motion (\ref{formula dotr3 dotp3}) and (\ref{intdotM}) are very complicated to be solved analytically. However, if we assume that $r_{kep}$ 
\begin{equation}
r_{kep}= \frac{p}{1+ e \cos \varphi}
\end{equation}
is the approximate solution of the restricted three-body problem, then by using numerical methods, we can integrate Eq.(\ref{intdotM}) approximately.   
where $e$ is the eccentricity, $p$ is the parameter of the orbit. 
In this manner, we can analyze the stability properties of test particles. Thus, we avoid the difficult problem of finding exact solutions of the three-body problem and, we analyze the stability properties of an entire class of approximate solutions. In this way, we expect to get some insight into the physical properties of the motion in the field of a three-body system.  

We acknowledge the support through a Grant of the Target Program of
the MES of the RK, Grant No. F. 0679, 2015.


\end{document}